\documentclass[a4paper,11pt]{article}
\usepackage{pos}
\usepackage{multirow}
\usepackage{cleveref}
\mathchardef\mhyphen="2D

\title{Heavy-light meson decay constants and hyperfine splittings with the heavy-HISQ method}
\ShortTitle{Heavy-light decay constants and hyperfine splittings}

\author*[a]{Kerr A. Miller}
\author[a]{Judd Harrison}
\author[a]{Christine T. H. Davies}
\author[b]{Antonio Smecca}

\affiliation[a]{School of Physics and Astronomy, University of Glasgow,\\
Glasgow, G12 8QQ, UK}

\affiliation[b]{Department of Physics, Swansea University,\\
Swansea, SA2 8PP, UK}

\emailAdd{k.miller.1@research.gla.ac.uk}
\emailAdd{judd.harrison@glasgow.ac.uk}
\emailAdd{christine.davies@glasgow.ac.uk}

\abstract{We compute ratios between the vector and pseudoscalar, and tensor and vector decay constants, and between hyperfine splittings for $D_{(s)}^{(*)}$ and $B_{(s)}^{(*)}$ mesons. We use the Highly Improved Staggered Quark (HISQ) action for all valence quarks, paired with the second generation MILC $n_f = 2+1+1$ HISQ gluon field configurations. These include light sea quarks with $m_u = m_d \equiv m_l$ going down to the physical values, as well as physically tuned strange and charm sea quarks. We also use a HISQ valence heavy quark, with mass ranging from that of the $c$-quark up to very nearly that of the physical $b$-quark on the finest lattices, allowing us to map out the heavy-quark mass dependence of the decay constant and hyperfine splitting ratios.}

\FullConference{The 41st International Symposium on Lattice Field Theory (LATTICE2024)\\
 28 July - 3 August 2024\\
Liverpool, UK\\}

\begin{document}
\maketitle

\section{Introduction}
\label{sec:intro}

Weak decays of mesons containing a heavy quark (i.e., bottom or charm) have recently been a source of much activity in the ongoing search for new physics \cite{Harrison:2020nrv, Harrison:2021tol, Gubernari:2022hxn, Harrison:2023dzh, FermilabLattice:2021cdg, Chakraborty:2021qav, Hurth:2020rzx, BESIII:2018hhz}. Phenomenological analyses of the leptonic decays of these mesons require precise knowledge of the relevant decay constants (DCs). While pseudoscalar DCs have been precisely determined using lattice QCD \cite{FlavourLatticeAveragingGroupFLAG:2024oxs}, lattice determinations of vector and tensor DCs are much less plentiful. In fact, as far as we are aware, there do not currently exist lattice calculations of the $B_{(s)}^*$ tensor DCs.

One precise method for determining these quantities is to compute the ratios of the vector and tensor DCs to the pseudoscalar DC, in which many correlated uncertainties cancel. One may then multiply these ratios by existing high-precision results for the pseudoscalar DC to determine the vector and tensor DCs with greater precision than might be attainable via a direct calculation.

In this work, we compute the vector-to-pseudoscalar and tensor-to-vector ratios of DCs for heavy-strange and heavy-light mesons with both bottom and charm quarks. For the heavy-light, we combine the heavy-strange ratio results with (double) ratios of heavy-strange ratios to heavy-light ratios, where -- again -- there is significant cancellation of correlated uncertainties. We use the heavy-HISQ method, varying our valence heavy-quark mass between that of the physical charm and physical bottom. This approach, originally applied with limited statistics to $B$ and $B_s$ decay constants in \cite{McNeile:2011ng}, has seen much recent success in the calculation of semileptonic form factors \cite{Parrott:2020vbe, Harrison:2023dzh, Harrison:2021tol, Harrison:2020gvo, Cooper:2020wnj}. We will make use of the recent fully non-perturbative calculation of the vector and tensor current renormalisation factors \cite{Hatton:2019gha, Hatton:2020vzp}. This calculation will also determine high-precision $B_{(s)}^{(*)}$ and $D_{(s)}^{(*)}$ masses, from which we compute a precise value for the heavy-strange hyperfine splitting, as well as the ratio of strange and light hyperfine splittings at both the $b$ and $c$ ends, enabling tests of the $\text{SU}(3)_\mathrm{flav}$ effects expected from heavy-quark effective theory (HQET) and chiral perturbation theory \cite{Jenkins:1992hx}.

\section{Theoretical Background}
\label{sec:theory}

In the continuum, decay constants may be expressed in terms of matrix elements of currents between the vacuum and suitable meson states. Here we consider states including an up/down or strange anti-quark and a heavy quark whose mass ranges between the charm and bottom quark masses. The interpolation/extrapolation of the heavy-quark mass to the physical $b$-quark mass will be discussed in \cref{sec:lattcalc}. The pseudoscalar decay constant of the $H_q$ state, and the vector and tensor decay constants of the $H_q^*$ state are defined via the relations
\begin{align}
\langle 0 | A_\mu | H_q(p)\rangle &\equiv p_\mu f_{H_q},\label{fPdef} \\
\langle 0 | V_\mu | H_q^*(p) \rangle &\equiv M_{H_q} f_{H_q^*} \epsilon_\mu(p),\label{fVdef} \\
\langle 0 | Z_T^{\overline{\mathrm{MS}}}\, T_{\alpha\beta} | H_q^*(p) \rangle &\equiv i f^T_{H_q^*} (\epsilon_\alpha p_\beta-\epsilon_\beta p_\alpha),\label{fTdef}
\end{align}
respectively, with $A_\mu = \bar{q} \gamma_\mu \gamma_5 h$, $V_\mu = \bar{q} \gamma_\mu h$, $T_{\alpha\beta} = \bar{q} \sigma_{\alpha\beta} h$, $h$ the valence heavy quark, and $\bar{q} = \bar{u}, \bar{d}, \bar{s}$ the valence light (anti-)quark. Note that the continuum tensor current includes a scheme-dependent renormalisation, for which we will use the $\overline{\mathrm{MS}}$ scheme.

\section{Lattice Calculation}
\label{sec:lattcalc}

We use the Highly Improved Staggered Quark~(HISQ) action for all valence quarks and the second generation MILC HISQ gauge configurations with $2+1+1$ sea quarks, including physical charm. Extending previous heavy-HISQ calculations, we use heavy-quark masses, $a m_h$, ranging from the tuned charm valence mass up to the physical $b$ valence mass on the three finest-lattice ensembles we have. Details of the configurations are given in~\cref{LattDesc1}. In the following, $q = l, s$ denotes the flavour of the lighter quark in the meson, where $l$ is the `light' (i.e., degenerate up/down) quark and $s$ is the strange quark.

\begin{table}
\centering
\begin{tabular}{c c c c c c c} 
\hline\hline
Set & $w_0/a$ & $N_x^3 \times N_t$ & $n_\mathrm{cfg} \times n_\mathrm{t}$ & $am_l^{\text{sea}}$ & $am_s^{\text{sea}}$ & $am_c^{\text{sea}}$ \\ [0.1ex] 
\hline
1  & 1.1119(10) & $16^3 \times 48$  & $1000 \times 16$ & 0.013   & 0.065   & 0.838 \\
\hline
2  & 1.1367(5)  & $32^3 \times 48$  & $1000 \times 16$ & 0.00235 & 0.00647 & 0.831 \\
\hline
3  & 1.3826(11) & $24^3 \times 64$  & $1000 \times 16$ & 0.0102  & 0.0509  & 0.635 \\
\hline
4  & \multirow{2}{*}{1.4149(6)} & \multirow{2}{*}{$48^3 \times 64$} & \multirow{2}{*}{$1000 \times 16$} & \multirow{2}{*}{0.00184} & \multirow{2}{*}{0.0507} & \multirow{2}{*}{0.628} \\
5  &            &                   &                  &         &         &       \\
\hline
6  & 1.9006(20) & $32^3 \times 96$  & $1000 \times 16$ & 0.0074  & 0.037   & 0.440 \\
\hline
7  & 1.9518(7)  & $64^3 \times 96$  & $614  \times  4$ & 0.0012  & 0.0363  & 0.432 \\
\hline
8  & 2.896(6)   & $48^3 \times 144$ & $500  \times 16$ & 0.0048  & 0.024   & 0.286 \\
\hline
9  & 3.0170(23) & $96^3 \times 192$ & $100  \times  4$ & 0.0008  & 0.022   & 0.260 \\
\hline
10 & 3.892(12)  & $64^3 \times 192$ & $375  \times  4$ & 0.00316 & 0.0158  & 0.188 \\
\hline\hline
\end{tabular}
\caption{Details of the 2+1+1 MILC HISQ gauge field configurations used in this study~\cite{MILC:2012znn,MILC:2010pul}. To fix the lattice spacing we use the Wilson flow parameter, which was found in~\cite{Dowdall:2013rya} to be $w_0=0.1715(9)$, together with the values of $w_0/a$~\cite{Chakraborty:2014aca,McLean:2019qcx} given in column 2. $n_\mathrm{cfg}$ and $n_\mathrm{t}$ in column 4 give the number of configurations and the number of time sources on each configuration respectively. $am_{l,\:\! s,\:\! c}^{\text{sea}}$ are the respective masses of the `light' (up/down), strange and charm sea quarks in lattice units.}
\label{LattDesc1}
\end{table}

\subsection{Current Renormalisation}
\label{sec:current-renorm}

In general, lattice operators are related to those in the continuum by renormalisation factors. Since we compute the pseudoscalar decay constant, $f_{H_q}$, using the partially conserved axial current relation, no renormalisation factor is required. For the vector operators, we use the local staggered current, the renormalisation factors for which, $Z_V$, were computed in \cite{Hatton:2019gha} in the RI-SMOM scheme. We also use the local staggered current for the tensor operators, with tensor renormalisation factors, $Z_T^c$, computed using an intermediate RI-SMOM scheme \cite{Hatton:2020vzp}. These were matched to the $\overline{\text{MS}}$ scheme at $\mu = 2\,\mathrm{GeV}$ and run to $\mu = 4.8\,\mathrm{GeV}$ using the corresponding 3-loop anomalous dimension \cite{Gracey:2000am}. In this work, we run these to $\mu = 0.9 \times M_{H_s}$, as a proxy for the heavy-quark pole mass.

\subsection{Correlation Functions}
Our correlation functions are constructed using the local staggered spin-taste operators $\gamma_5 \otimes \gamma_5$ (pseudoscalar), $\gamma_i \otimes \gamma_i$ (vector) and $\gamma_i \gamma_t \otimes \gamma_i \gamma_t$ (tensor), with the vector and tensor correlators averaged over the spatial directions. For simplicity, we express the correlation functions in terms of the analogous Dirac-spinor operators. We construct the 2-point correlation functions
\begin{equation}\label{CJ}
    C_J (t,0) = \left\langle \bar{q}\, \Gamma^J h(t)\, \bar{h} \Gamma^J q(0) \right\rangle = \sum_n \left( \left| A^J_n \right|^2 e^{-t M^J_n} - (-1)^t \left| \tilde{A}^{J}_n \right|^2 e^{-t \tilde{M}_n^{J}} \right),
\end{equation}
where $\Gamma^{P,V,T} = \gamma^5,\, \gamma^i,\, \gamma^i \gamma^t$ corresponds to the pseudoscalar, vector or tensor current respectively. The operator $\bar{q}\, \Gamma^J h(t)$ is projected to zero momentum and we use a random wall source for $\bar{h} \Gamma^J q(0)$ to increase statistics. The right-hand side of~\cref{CJ} is the spectral decomposition of the 2-point function, using a complete set of states. Time-doubled states produce the time-oscillating terms, a generic feature of using staggered quarks~\cite{Follana:2006rc}.

The ground-state, non-oscillating amplitudes, $A^J_0$, which we extract from our fits, are related to the decay constants by
\begin{alignat}{2}
A^P_0 &= \frac{\left\langle 0 \middle| \bar{q}\gamma^5 h \middle| H_{q} \right\rangle}{ \sqrt{2 M_{H_{q}}} } &&= \frac{ M_{H_{q}}^{3/2} }{\sqrt{2}\, (m_h + m_q)} f_{H_{q}}, \label{fPeq} \\
A^V_0 &= \frac{\left\langle 0 \middle| \bar{q}\gamma^i h \middle| H_{q}^* \right\rangle}{ \sqrt{2 M_{H_{q}^*}} } &&= \frac{ \sqrt{M_{H_{q}^*}} }{\sqrt{2}\, Z_V} f_{H_{q}^*}, \label{fVeq} \\
A^T_0 &= \frac{\left\langle 0 \middle| \bar{q} \gamma^i \gamma^t h \middle| H_{q}^* \right\rangle}{ \sqrt{2 M_{H_{q}^*}} } &&= \frac{ \sqrt{M_{H_{q}^*}} }{\sqrt{2}\, Z^c_T} f^T_{H_{q}^*}. \label{fTeq}
\end{alignat}
We also construct 2-point functions for the flavour-diagonal pseudoscalars $\pi\, (l\bar{l}\,)$, $\eta_s\, (s\bar{s})$ and $\eta_c\, (c\bar{c})$, generically denoted $\eta_q$, using the local $\gamma_5 \otimes \gamma_5$ operator (which does not receive contributions from oscillating states in the flavour-diagonal case). We fit these to the form
\begin{equation}\label{Ceta}
C_{\eta_q}(t,0) = \sum_n \left| A^{\eta_q}_n \right|^2 e^{-t M^{\eta_q}_n}.
\end{equation}

\subsection{Continuum Extrapolation}

To analyse our lattice data, we perform fully simultaneous fits, including all correlations, using the \texttt{corrfitter} python package \cite{corrfitter}. In the following, we describe the extrapolation of our lattice data to the physical $m_h = m_b$ point. We work with the ratios of decay constants, as discussed in~\cref{sec:intro}, first focusing on the $q = s$ case: $f_{H_{s}^*}^T \big/ f_{H_{s}^*}$ and $f_{H_{s}^*} / f_{H_{s}}$ in~\cref{decayconstextrapsec}, and the hyperfine splitting, $\Delta_{H_s^*-H_s}\equiv M_{H_s^*}-M_{H_s}$, in~\cref{hyperfinesec}. We then give the ratios of these quantities between the $q = s$ and $q = l$ cases in~\cref{sec:double-ratios-and-hfs-ratio}.

\subsubsection{Decay Constant Ratios}
\label{decayconstextrapsec}

\begin{figure}
\centering
\includegraphics[scale=0.35]{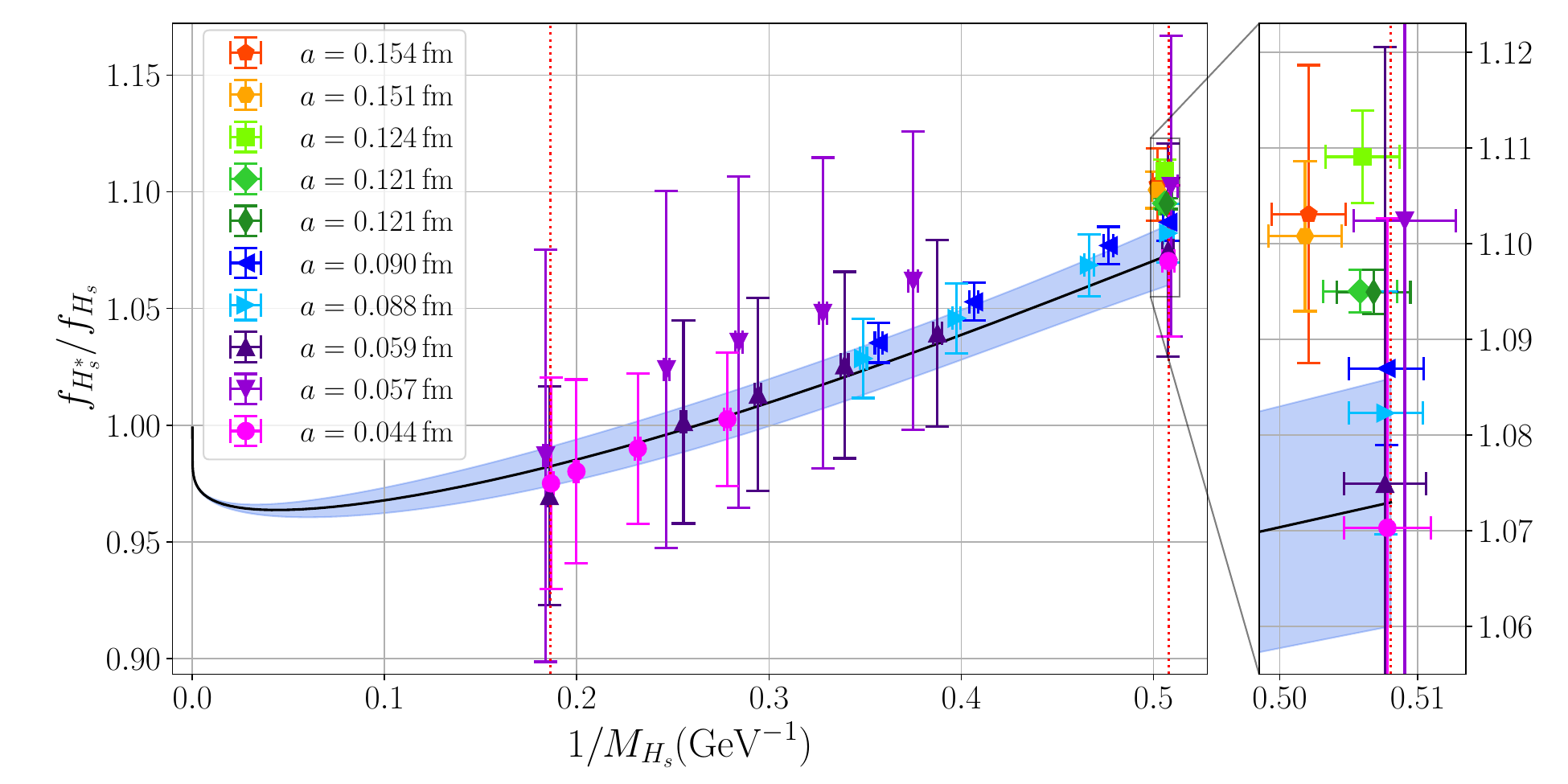}
\includegraphics[scale=0.35]{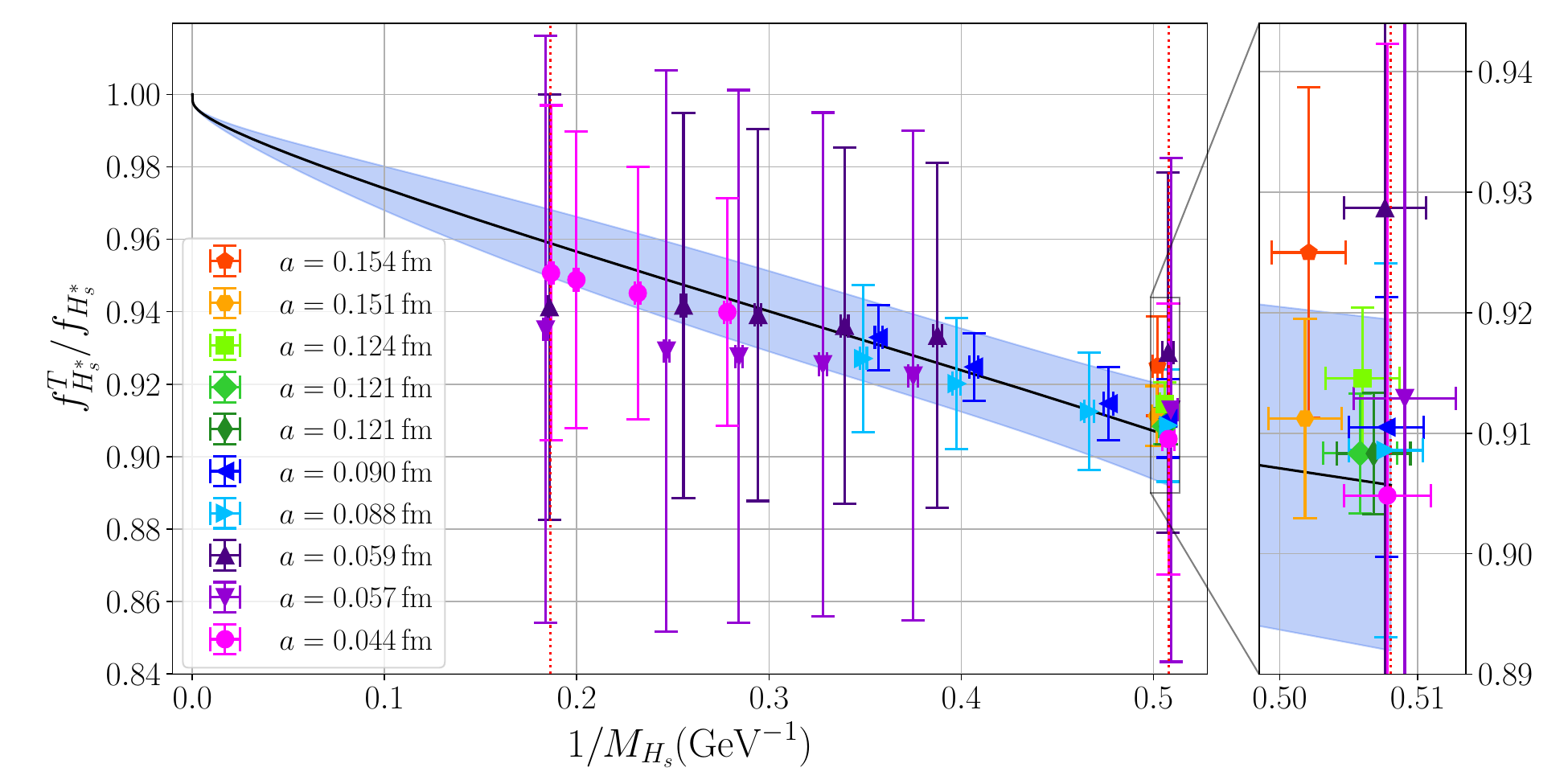}
\caption{The ratios $f_{H_s^*} / f_{H_s}$ (top) and $f^T_{H_s^*} \big/ f_{H_s^*}$ (bottom) plotted against $1/M_{H_s}$. The blue bands and black curves are our physical continuum results. The physical $B_s$ and $D_s$ masses correspond to the left and right red, dotted, vertical lines, respectively. The panels on the right show zoomed-in views of the plots around the $D_s$ physical point. \label{DCratios}}
\end{figure}

To reach the physical point, we fit our lattice data to a form designed to capture discretisation and quark mistuning effects, in addition to physical heavy-quark mass dependence and analytic chiral dependence. Following \cite{McNeile:2011ng}, we use a form inspired by the leading $m_h$ dependence of the decay constant in HQET \cite{Buchalla:2002pd}. We denote the ratio being considered ${R_s}$, where $R_{s}\!~=~\!\left( f_{H_s^*}^T \middle/ f_{H_s^*} \right),\, \left( f_{H_s^*} \middle/ f_{H_s} \right)$, and use the fit function
\begin{equation}
    R_{s} = 1 + \sum_{i=1}^4 c^{R_s}_{i} \left( \frac{{\alpha_s}{\left( \hat{m}_h \right)}}{\pi} \right)^i
    + \mathcal{N}^{R_{s}} \sum^3_{i,j,k = 0} C^{R_s}_{ijk} \left( \frac{\Lambda_{\text{QCD}}}{M_{H_{s}}} \right)^{i}
    \left( \frac{am_h}{\pi} \right)^{2 j}%
    \left( \frac{M_{K}^2}{\Lambda_{\chi}^2} \right)^{k},
    \label{eq:DC-ratio-fit}
\end{equation}
where $c_i^{R_s}$ are the third-order perturbative coefficients computed in \cite{Bekavac:2009zc}. For $i\geq 2$ these coefficients are nontrivial functions of $m_c^\mathrm{pole,\, sea} \big/ m_h^\mathrm{pole}$. As in~\cref{sec:current-renorm}, we approximate the pole mass using $m_h^\mathrm{pole} = 0.9 \times M_{H_s}$. For the charm quark pole mass, we use $m_c^\mathrm{pole,\, sea} = 0.9 \times (M_{D_s} + m_c^\mathrm{sea} - m_c^\mathrm{val})$, correcting for the difference between sea and valence masses. The factor $\mathcal{N}^{R_{s}}$ includes sea and valence quark mass mistuning dependence, given by
\begin{align}
    \mathcal{N}^{R_{s}} &= 1 + A \delta_{m_l}^\mathrm{sea} + B \delta_{m_s}^\mathrm{sea} + C \delta_{m_c}^\mathrm{sea} + D \delta_{m_s}^\mathrm{val}, \label{quark-mistuning-N} 
\end{align}
\begin{align}
    \begin{split}
    \delta_{m_c}^\text{sea} &= (am_{c}^\text{sea} - am_c^\text{tuned}) \big/ am_c^\text{tuned},\\
    \delta_{m_s}^\text{val} &= (am_{s}^\text{val} - am_{s}^\text{tuned}) \big/ (10 am_{s}^\text{tuned}),\\
    \delta_{m_{s(l)}}^\text{sea} &= (am_{s(l)}^\text{sea} - am_{s(l)}^\text{tuned}) \big/ (10 am_{s}^\text{tuned}).
    \end{split}
\label{deltatermseq}
\end{align}
The above tuned quark masses are given by
\begin{equation}
a m_l^\text{tuned} = a m_s^\text{tuned} \big/ [m_s / m_l]^\mathrm{phys},\quad
a m_{s(c)}^\text{tuned} = a m_{s(c)}^\text{val} \left( \frac{M_{\eta_s (\eta_c)}^\text{phys}}{M_{\eta_s (\eta_c)}} \right)^{2(1.4)},
\end{equation}
with $[m_s / m_l]^\mathrm{phys} = 27.18(10)$ \cite{Bazavov:2017lyh} and $M_{\eta_c}^\mathrm{phys}$ the QCD-only, quark-line connected value determined in \cite{Hatton:2020qhk}. We use $\alpha_{\overline{\mathrm{MS}}}(5\,\mathrm{GeV}, n_f = 4) = 0.2128(25)$ \cite{Chakraborty:2014aca}, which we run to scale $\mu$ using the 4-loop results for the beta function \cite{vanRitbergen:1997va} and convert to the $V$ scheme using the expressions in \cite{Schroder:1998vy, Lepage:1992xa}. We do not include a $\delta_{m_c}^\mathrm{val}$ term in \cref{quark-mistuning-N}, as this dependence is captured in the physical heavy mass dependence term $({\Lambda_\mathrm{QCD}}/{M_{H_s}})^{i}$. Our continuum extrapolated decay constant ratios for the $q = s$ case, $f_{H_s^*} / f_{H_s}$ and $f^T_{H_s^*} \big/ f_{H_s^*}$, are plotted against $1/M_{H_s}$ in \cref{DCratios}.

\subsubsection{Hyperfine Splittings}
\label{hyperfinesec}

\begin{figure}
\centering
\includegraphics[scale=0.35]{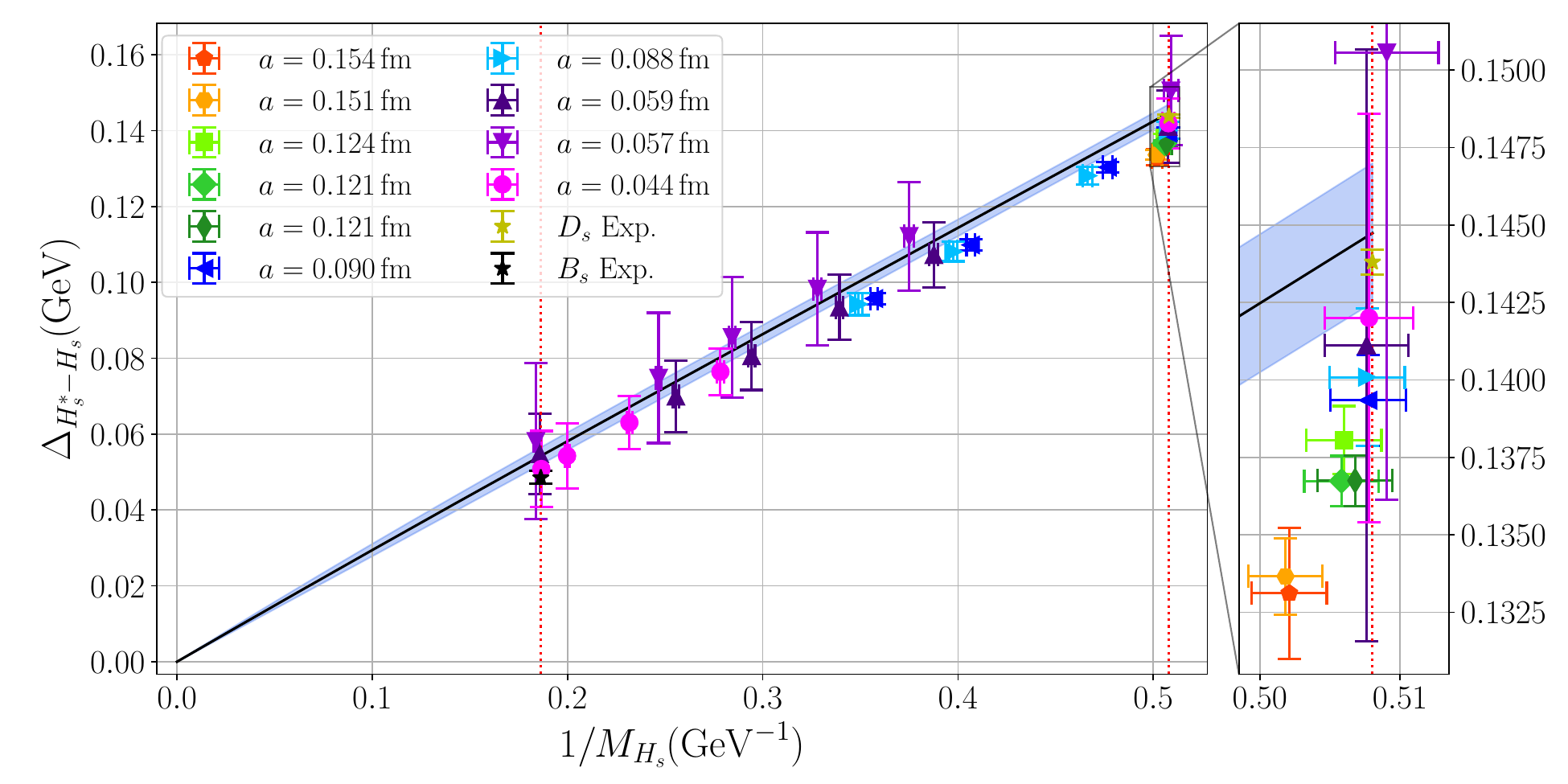}
\caption{The hyperfine splitting $\Delta_{H_s^* - H_s}\equiv M_{H_s^*} - M_{H_s}$ plotted against $1/M_{H_s}$. The blue band and black curve are our physical continuum result. The physical $B_s$ and $D_s$ masses correspond to the left and right red, dotted, vertical lines, respectively. The experimental values of $\Delta_{B_s^* - B_s}$ and $\Delta_{D_s^* - D_s}$ are also plotted for comparison. The panel on the right shows a zoomed-in view of the plot around the $D_s$ physical point. \label{splittingsplots}}
\end{figure}

Our fit results for the meson masses also allow us to compute the phenomenologically interesting hyperfine splittings, $\Delta_{H_s^* - H_s} = M_{H_s^*} - M_{H_s}$. We use a similar fit function to \cref{eq:DC-ratio-fit}:
\begin{align}\label{fitfunctioneqhyp}
\Delta_{H_s^* - H_s} = \mathcal{N} \sum_{i,j,k = 0}^3 C_{ijk}
\left(\frac{\Lambda_\mathrm{QCD}}{M_{H_s}}\right)^i
\left( \frac{a m_h}{\pi} \right)^{2j}
\left( \frac{M_K^2}{\Lambda_\chi^2} \right)^{k},
\end{align}
where $\mathcal{N}$ has the same form as \cref{quark-mistuning-N}, and we set $C_{00k} = 0$, corresponding to the expected static limit in the continuum: $\Delta_{H_s^* - H_s} \to 0$ as $m_h \to \infty$. The result of this extrapolation is shown in \cref{splittingsplots}, together with our lattice data points.

\subsubsection{Decay Constant Double Ratios and Hyperfine Splitting Ratio}
\label{sec:double-ratios-and-hfs-ratio}

\begin{figure}
\centering
\includegraphics[scale=0.35]{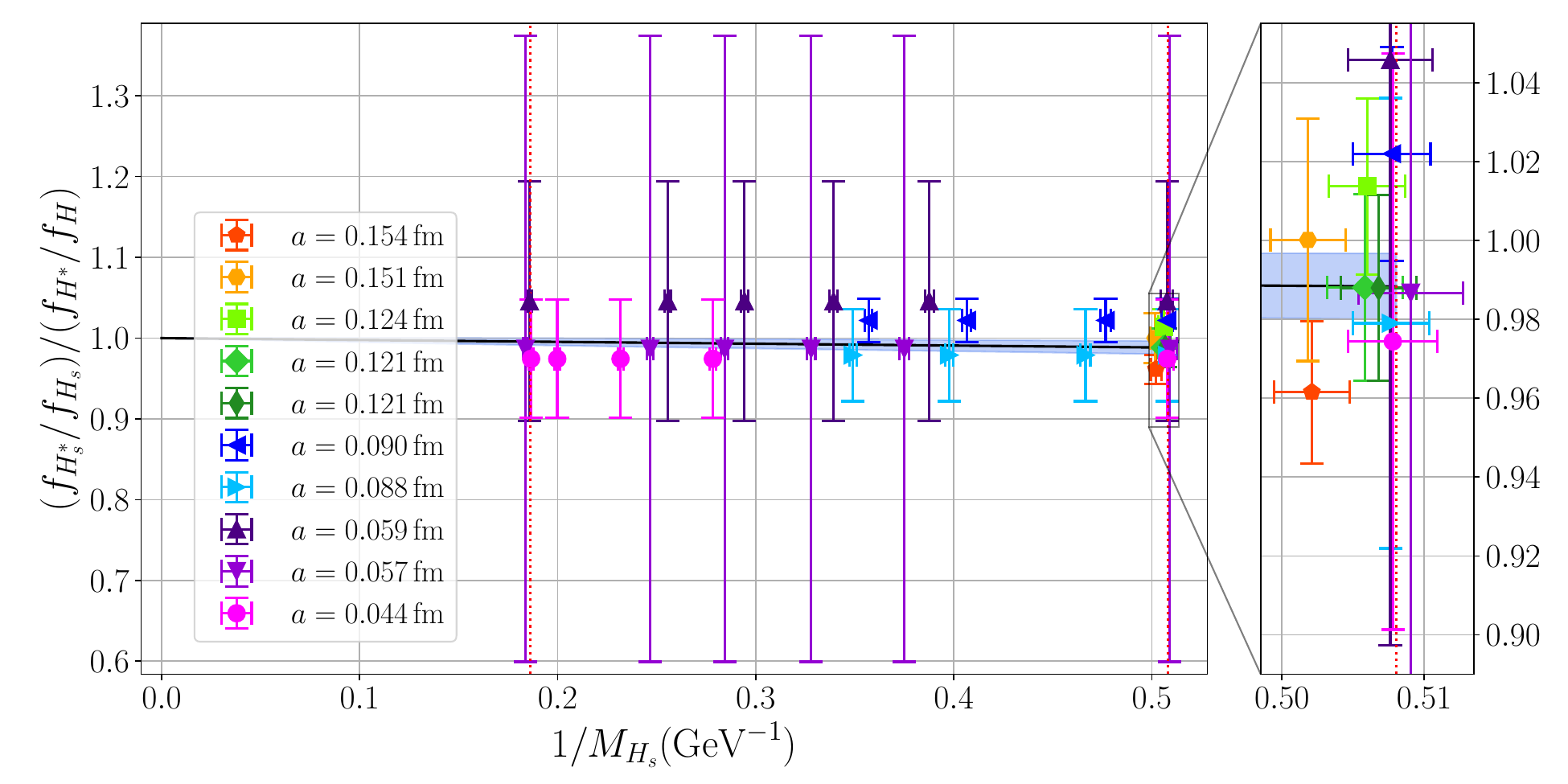}\\
\includegraphics[scale=0.35]{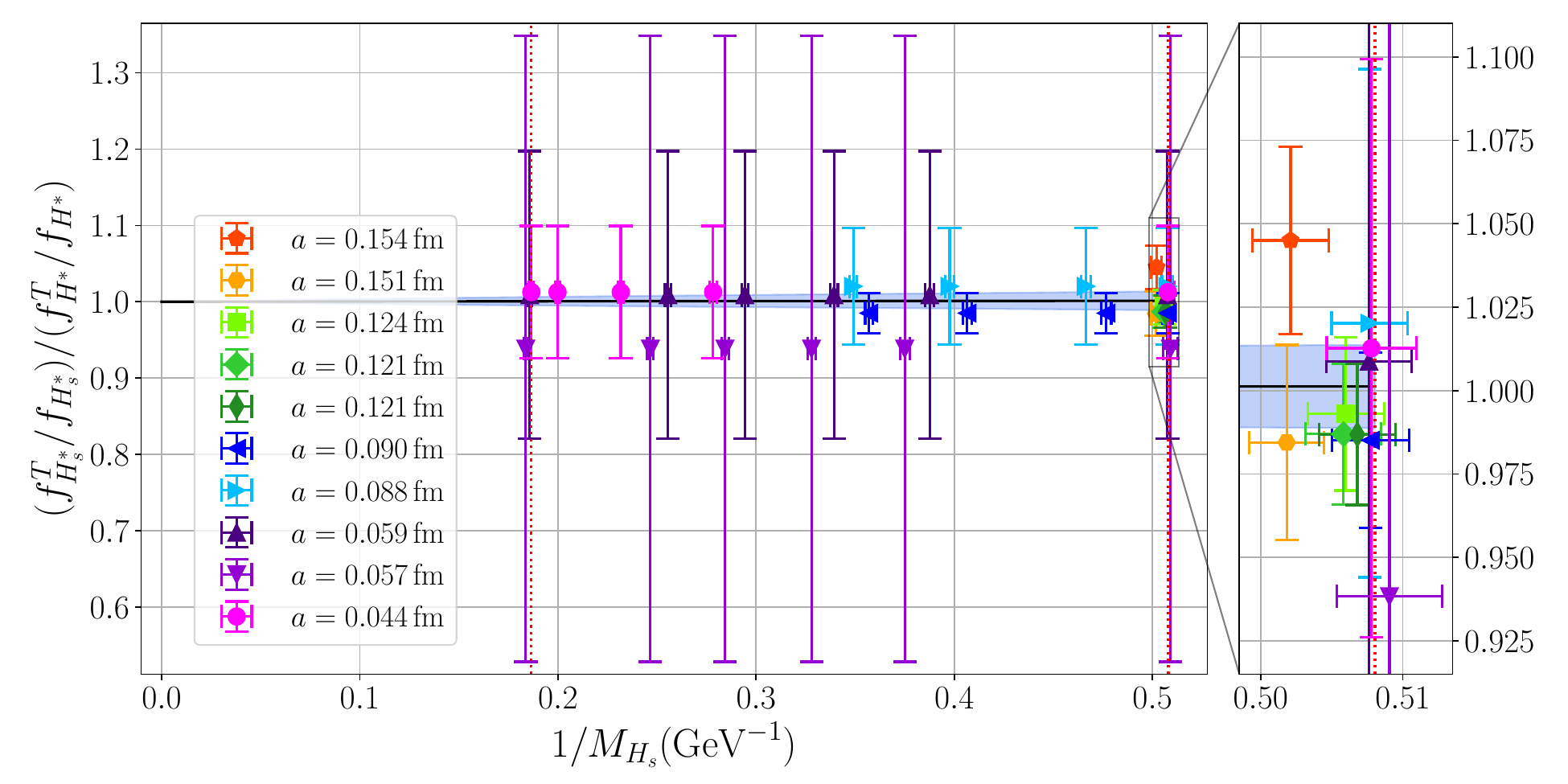}
\caption{The double ratios $\frac{ f_{H_s^*} / f_{H_s}}{ f_{H^*} / f_{H}}$ (top) and $\frac{ f_{H_s^*}^{T} \big/ f_{H_s^{*}}}{ f_{H^*}^{T} \big/ f_{H^{*}}}$ (bottom) plotted against $1/M_{H_s}$. The blue bands and black curves are our physical continuum results. The physical $B_s$ and $D_s$ masses correspond to the left and right red, dotted, vertical lines, respectively. The panels on the right show zoomed-in views of the plots around the $D_s$ physical point. \label{doubleratioplots}}
\end{figure}

\begin{figure}
\centering
\includegraphics[scale=0.35]{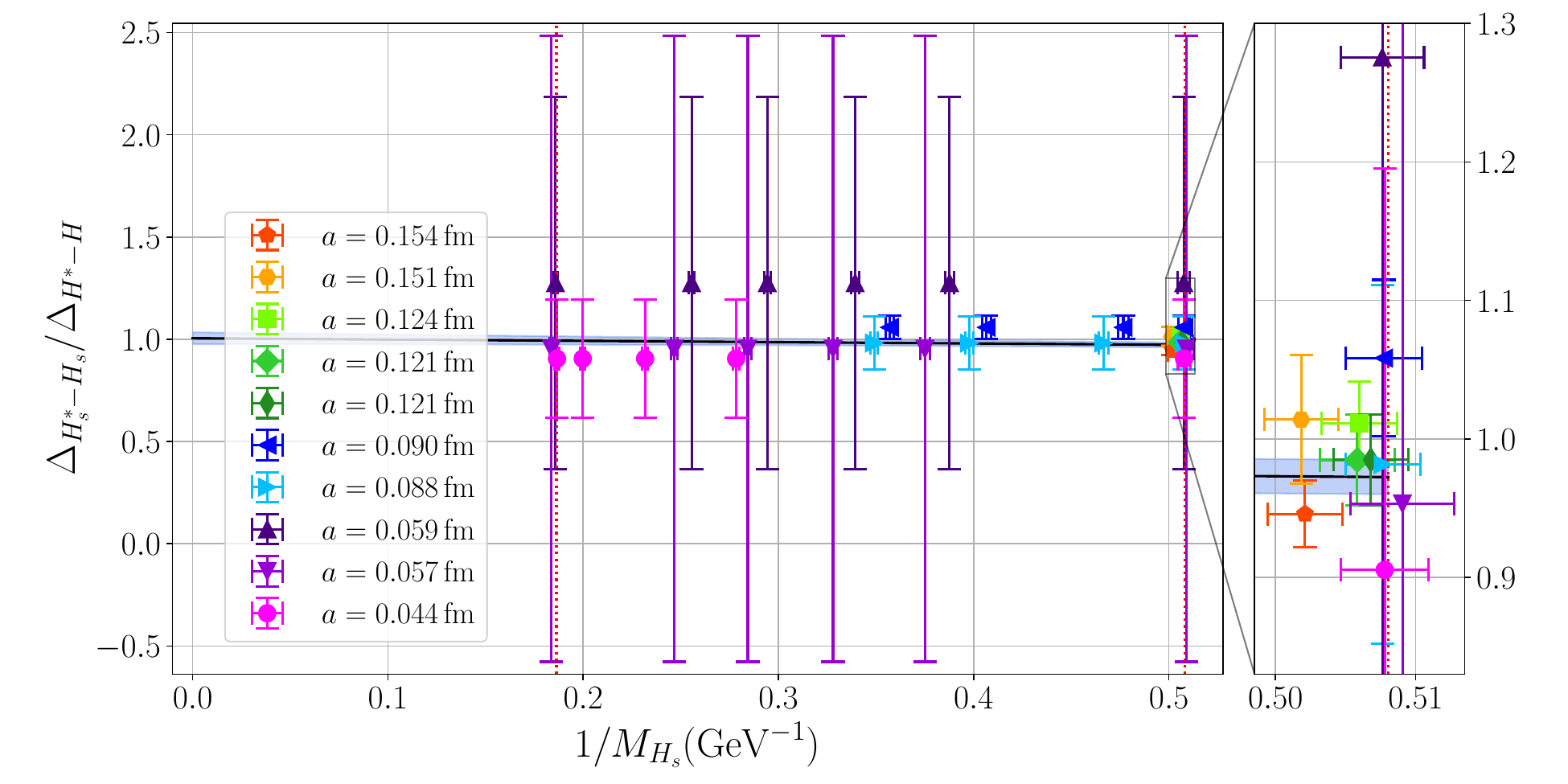}
\caption{The ratio of hyperfine splittings $\frac{\Delta_{H_s^* - H_s} }{ \Delta_{H^* - H}}$ plotted against $1/M_{H_s}$. The blue band and black curve are our physical continuum result. The physical $B_s$ and $D_s$ masses correspond to the left and right red, dotted, vertical lines, respectively. The panel on the right shows a zoomed-in view of the plot around the $D_s$ physical point. \label{hyperfineratioplot}}
\end{figure}

A similar fit form was used for the double ratios of decay constants (i.e., ratios of the heavy-strange ratios to the heavy-light ratios) and the ratio of the heavy-strange and heavy-light hyperfine splittings. We denote the (double) ratio being fitted by $X$, where 
\begin{equation*}
X \in \left\{ \frac{ f_{H_s^*} / f_{H_s} }{ f_{H^*} / f_{H} },\, \frac{ f_{H_s^*}^T \big/ f_{H_s^*} }{ f_{H^*}^T \big/ f_{H^*} },\, \frac{ \Delta_{H_s^* - H_s} }{ \Delta_{H^* - H} } \right\}.
\end{equation*}
The fit form is then
\begin{equation}
    X = 1 + \mathcal{N}^X \left( \frac{M_K^2 - M_{\pi}^2}{\Lambda_{\chi}^2} \right) 
    \sum_{i, j = 0}^3 C^X_{ij} \left( \frac{\Lambda_{\text{QCD}}}{M_{H_{s}}} \right)^i%
    \left( \frac{am_h}{\pi} \right)^{2j},
    \label{eq:dr-and-hfs-ratio-fit}
\end{equation}
where, again, $\mathcal{N}^X$ has the same form as \cref{quark-mistuning-N}, and the overall factor of $(M_K^2 - M_{\pi}^2)/\Lambda_\chi^2$ sets the size of $\text{SU}(3)_\mathrm{flav}$-breaking effects. For the double ratios of decay constants, we fix $C_{00} = 0$ to ensure the correct limit as $m_h \to \infty$. The double ratios are shown in \cref{doubleratioplots}, where we see that our results for both ratios are consistent with 1 for $m_h \geq m_c$, and the ratio of hyperfine splittings is shown in \cref{hyperfineratioplot}.

\section{Conclusions and outlook}
\label{sec:conclusions}

Using the heavy-HISQ method, we have determined high-precision ratios between the vector and pseudoscalar decay constants, and between the tensor and vector decay constants of $B_s^{(*)}$ and $D_s^{(*)}$ mesons, as well as ratios of these ratios between the strange and `light' (up/down) light-flavour cases. Similarly, we have computed $B_s^{(*)}$ and $D_s^{(*)}$ hyperfine splittings and ratios of these quantities to those of their heavy-light counterparts. By combining our results with the high-precision pseudoscalar calculations of \cite{Bazavov:2017lyh}, we will obtain precise values for the vector and tensor decay constants of the $B^*, B_s^*, D^*$ and $D_s^*$ mesons, some of which will be the first ever lattice results. These quantities enable precision tests of Standard-Model heavy-flavour physics and constitute vital inputs to future theoretical work, including lattice calculations of phenomenologically interesting semileptonic decay processes. To complete this work, we will also finalise testing the stability of our fitting procedure, incorporate QED corrections to the hyperfine splittings, and calculate the phenomenological implications of our results.

\acknowledgments

We are grateful to the MILC Collaboration for the use of
their configurations and code. This work used the DiRAC
Data Intensive service (CSD3) at the University of
Cambridge, managed by the University of Cambridge
University Information Services on behalf of the STFC
DiRAC HPC Facility (www.dirac.ac.uk). The DiRAC
component of CSD3 at Cambridge was funded by BEIS, UKRI
and STFC capital funding (e.g., Grants No. ST/P002307/1
and No. ST/R002452/1) and STFC operations grants (e.g.,
Grant No. ST/R00689X/1). DiRAC is part of the UKRI
Digital Research Infrastructure. We are grateful to the
CSD3 support staff for assistance. Funding for this work
came from the University of Glasgow LKAS Scholarship
fund, UK STFC Grants No. ST/L000466/1, No. ST/P000746/1,
No. ST/X000605/1, and No. ST/T000945/1 and EPSRC Project
No. EP/W005395/1.

\bibliographystyle{JHEP_mod}
\bibliography{Lattice24PoS}

\end{document}